\documentclass[prl,aps,twocolumn]{revtex4-1} 
\usepackage[normalem]{ulem}
\usepackage{graphicx} 
\usepackage[usenames]{color}
\usepackage{amsmath,amssymb}
\usepackage{wasysym}
\usepackage{gensymb}
\usepackage{bbold}
\usepackage{natbib}

\usepackage[usenames,dvipsnames]{xcolor}
\usepackage{lipsum}

\usepackage{color}
\definecolor{bleu}{rgb}{0.1,0.2,0.6}

\newcommand{\nsub}{n_\text{sub}}

\newcommand{\eg}{{\em e.g.} }

\begin{document} 
\title{Mixing by Unstirring: Hyperuniform Dispersion of Interacting Particles upon Chaotic Advection}
\author{Joost H. Weijs}
\affiliation{Univ Lyon, Ens de Lyon, Univ Claude Bernard, CNRS, Laboratoire de Physique, F-69342 Lyon, France}

\author{Denis Bartolo}
\affiliation{Univ Lyon, Ens de Lyon, Univ Claude Bernard, CNRS, Laboratoire de Physique, F-69342 Lyon, France} 
\date{\today} 

\begin{abstract}
We show how to achieve both fast and hyperuniform dispersions of particles in viscous fluids.
To do so, we first extend the concept of critical random organization to chaotic drives. We  show how palindromic sequences of chaotic advection cause microscopic particles to effectively interact at long range thereby inhibiting critical self-organization. Based on this understanding we go around this limitation and design sequences of stirring and unstirring which simultaneously optimize the speed of particle spreading and the homogeneity of the resulting dispersions.

\end{abstract} 
\maketitle

Mixing a concentrated batch of particles  through a viscous fluid is
 a challenging task common to a host of industrial and everyday processes. 
Take for example the mixing of chocolate chips through cookie dough: ideally one would want to mix them as uniformly as possible, such that each cookie ends up with the very same amount of chips without having to place them one at  a time. Beyond culinary applications, from the production of concrete to foundation cream,  the challenge is always twofold: dispersing the particles throughout the entire sample as fast {\em and} as homogeneously as possible.

Fast mixing can be achieved  even in viscous pastes  where turbulence and diffusion are ineffective.  The most common strategy is known as chaotic advection~\cite{aref1984,OttinoARFM1990,StroockARCB2012}, and has been used by cookie makers for centuries. It consists in stirring the fluid by repeated sequences stretching and folding.
{\color{black}However, in incompressible fluids, chaotic advection cannot yield number fluctuations smaller than that of a random set of non-interacting particles \cite{DuringPRE2009}.}  The fluctuations of the number of particles, $N(\ell),$ in a region of size $\ell$ is bound to scale as  its mean value $\Delta N^2(\ell) \sim \langle N(\ell)\rangle$.

In contrast, nearly perfect homogenization methods have been  proposed in a very different context. Numerical simulations have shown how, when periodically driven, ensembles of  interacting particles undergo a critical transition from a time-reversible to an irreversible dynamics. The time-reversible state corresponds to an absorbing state where the strobed dynamics  of the particles is frozen~\cite{CorteNature2008,JeanneretNatCom2014,WeijsPRL2015,KeimPRL2014,Okuma2011}. This phenomenon, coined random organization, was first demonstrated for periodically sheared suspensions~\cite{PineNature2005,CorteNature2008,KeimPRL2014}, and subsequently reported for a number of physical systems ranging from emulsions~\cite{JeanneretNatCom2014,WeijsPRL2015} to granular media \cite{Slotterback2012,RoyerPNAS2015} to driven vortices in superconductors \cite{Okuma2011}.
At the critical point, numerical simulations predict  the  self-organization of the particles into amorphous hyperuniform structures~\cite{LevinePRL2015,Berthier2015,WeijsPRL2015,Tjhung2016,HexnerPRL2017}. 
A hyperuniform set is defined as a ensemble of objects with reduced number fluctuations: $\Delta N^2/\langle N\rangle\sim \ell^\lambda$, with $-1<\lambda<0$ (\eg $\lambda=-1$ for a perfect periodic lattice)~\cite{TorquatoPRE2003}. 
However, hyperuniformity comes at a price in periodically driven systems: the underlying self-organization dynamics is intrinsically very slow. Firstly, it is subject to critical slowing down, secondly and even more importantly, all the available models are inefficient at dispersing ensembles of particles initially concentrated in compact regions.

In this letter, we demonstrate numerically how to achieve fast {\em and} hyperuniform mixing.   
We first establish that chaotic advection and random organization are intrinsically incompatible in finite-size systems.
{\color{black} Using a prototypical model of {\em interacting} particles driven by a time-periodic chaotic flow}, we show how criticality and thus hyperuniform ordering are suppressed by long-range effective interactions caused by the sequential stretching and folding sequences intrinsic to chaotic flows. 
Finally, we show how to go around this incompatibility and design time-dependent flows which practically combine the speed of chaotic advection and achieve the hyperuniform fluctuations of random organization. 
\begin{figure}
\begin{center}
\includegraphics{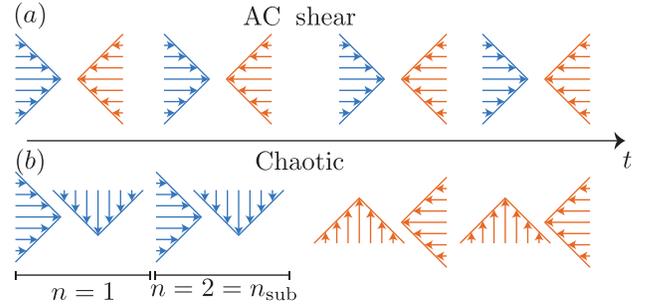}
  \caption{
  \label{fig:trajectories} 
  Sketch of the two palindromic  flow fields. 
  (a)  AC shear cycles, consisting of a forward stirring phase (blue) mirrored by the time-reversed unstirring phase (orange). Four cycles are shown.
  (b) A single palindromic sequence that results in chaotic flow. Each cycle  consists of one stirring phase (blue) followed by the time-reversed unstirring phase  (orange). Both phase are here composed
  of $\nsub=2$ subcycles made of alternating horizontal and vertical shear flows.
  }
  \end{center}
\end{figure}
\begin{figure*}[t]
\begin{center}
  \includegraphics{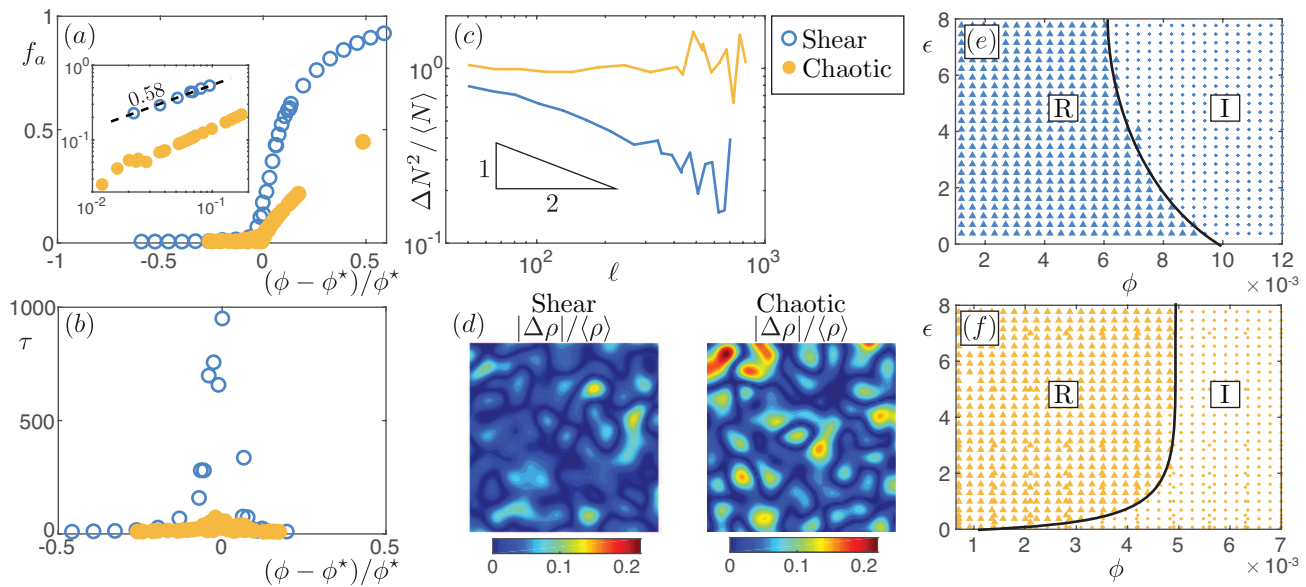}
  \caption{
  \label{fig:second!} 
  Comparison between pure shear ($t_x=1$) and chaotic  driving flow ($t_x=0.5$).
  (a) Steady state active fraction $f_a$ at various densities $\phi$.
  Inset log-log scale.
  (b) Relaxation time $\tau$ of the active fraction $f_a(t)$, as fitted by $f_a(t)=(f_a(0)-f_a(\infty)) \frac{\exp(-t/\tau)}{t^\delta}+f_a(\infty)$~\cite{CorteNature2008}. 
$\tau$ diverges according to a power-law $\tau\sim |\phi-\phi^\star|^{1.2\pm 0.2}$, see also~\cite{Supp}. 
This behaviour is not seen for the chaotic driving. 
  (c) Number  fluctuations at varying scale $\ell$ in the steady state for   AC shear and palindromic chaotic drive with $n_{\rm sub}=10$, $\gamma_0=3$. 
  A decrease of $\left(\left<N^2\right>-\left<N\right>^2\right)/\left<N\right>$ with $\ell$ for all $\ell$s indicates a fully developed hyperuniform structure. 
Under pure shear hyperuniformity clearly develops around $\phi^\star$ whereas it is absent under chaotic driving. 
  Note that this quantity has been corrected by a factor $1/(1-\ell^2/L^2)$, to compensate for a reduction in number fluctuations as $\ell\rightarrow L$ due to the conserved total number of particles in the finite-size system \cite{TrizacEJP2014}.
  (d) {\color{black} Instantaneous normalized density field $\left|\rho-\langle\rho\rangle\right|/\langle\rho\rangle$ for both simulations, demonstrating the reduced number fluctuations for the pure-shear case. $\rho(\mathbf r,t)$ is defined as the number of particles in regions of size $3.5a$.}
  (e,f) Phase diagrams ($\phi,\epsilon$) showing the reversible and irreversible regimes for the simple shear (e) and the chaotic flow (f). 
}
\end{center}
\end{figure*}

{\color{black}We consider a model which generalizes the random-organization model introduced in ~\cite{CorteNature2008}.  $N$ particles of diameter $a$  are advected by a time-dependent flow and interact at contact. Between two contact events the particle positions evolves according to $\partial_t\mathbf r_i(t)=\mathbf v(\mathbf r_i(t),t)$, where $i=1\ldots N$, and $\mathbf v(\mathbf r,t)$ is a time dependent flow field (we discuss the range of validity of this instantaneous-response approximation in \cite{Supp}).  Following~\cite{CorteNature2008} we model the particle collisions with random kicks upon contact, and note $\epsilon$ the maximum amplitude of the random kicks. The parameter $\epsilon$ accounts for microscopic details of the contact interactions such as particle roughness~\cite{DavisPOFA1992,dacunhaJFM1996,Metzger_PRFl_2015}. This simplified interaction has proven to yield very good agreement with full Stokesian dynamics simulations including contact interactions~\cite{PineNature2005,MetzgerPRE2013,MetzgerPOF2015}. 
In all that follows we focus on 2D systems and use square boxes with periodic boundary conditions. We stress that due to collisions  particles do not behave as passive tracers of the flow and can self-organize. }
%

The flow field used in our numerical simulations is a continuous-time version of the tent map that alternates between shears of amplitude $\gamma_0$ along the horizontal ($+x$) and vertical ($-y$) directions, it is defined as follows:
\begin{align}
\label{eq:flowfield}
{\bf v}(\mathbf r,t)=&\gamma_0 (1-|2y-1|) \hat{\bf x} &&\text{for }0  <t \;({\rm mod}\;1)<t_x\nonumber\\
{\bf v}(\mathbf r, t)=&\gamma_0 (|2x-1|-1) \hat{\bf y} &&\text{for }t_x<t \;({\rm mod}\;1)<1,
\end{align}
where $\hat {\mathbf x}$ (resp. $\hat {\mathbf y}$) is the unit vector pointing along the $x$-direction (resp. $y$-direction). Here  $x,y \in [0,1]$, with the simulation box size normalized to $1$.  The two type of flows we consider are distinguished by the value of $t_x$.
When $t_x=1$ the flow reduces to a horizontal shear flow, Fig.~\ref{fig:trajectories}a. Instead when $t_x=\frac 1 2$ alternating horizontal and vertical shears are applied, Fig.~\ref{fig:trajectories}b, which yield chaotic advection as exemplified in~\cite{Supp}.{ \color{black}We  investigate the impact of unstirring on particle organization, by considering flows composed of palindromic sequences.  As illustrated in Fig.~\ref{fig:trajectories}, the stirring phase of each palindrome is composed of $n_{\rm sub}$ units  termed subcycles and defined by Eq.~\eqref{eq:flowfield}. The palindrome is completed by the corresponding unstirring pattern. Fig.~\ref{fig:trajectories}  compares an AC shear flow to a   palindromic   sequence with $n_{\rm sub}=2$.}

One technical comment is in order: in silico, even in the absence of collisions the chaotic nature of the flow prevents reversible trajectories when using a floating point representation of their position. We therefore use a custom fine integer discretization thoroughly described in~\cite{Supp}.
This method ensures that $\epsilon$ is the only source of irreversibility in the simulations.

Let us first investigate the impact of Lagrangian chaos on the nature of the reversible-to-irreversible transition, and on the resulting spatial structures. We compare the strobed dynamics of two systems driven by two different palindromic  sequences of shear flows, starting from fully dispersed random initial conditions. The first flow  is a simple AC shear, and corresponds to $t_x=1$. The second flow corresponds to $t_x=\frac1 2$ and $n_{\rm sub}=10$ which results in a sequence of stirring and unstirring that is highly chaotic, Supplementary Videos 1 and 2. The state of reversibility of the particle dynamics is measured by the fraction of particles {\em not} returning to their initial position at the end of the palindromic sequence and henceforth called active particles~\cite{CorteNature2008}. 

In Fig.~\ref{fig:second!}(a), the active fraction  in  steady-state, $f_a$, is plotted as a function of particle area fraction $\phi$, where $f_a=0$ denotes a fully reversible state.
For both   AC shear  and  chaotic flows, we clearly observe a transition from a reversible ($\phi<\phi^\star$) to an irreversible dynamics ($\phi>\phi^\star$).
In the case of simple AC shear, we find that $f_{\rm a}\sim(\phi-\phi^\star)^\beta$, where $\beta=0.58$ and that the relaxation time of this order parameter diverges at the transition as $\tau\sim(\phi-\phi^\star)^{z\nu}$, where $z\nu\approx1.2$, see Figs.~\ref{fig:second!}a and ~\ref{fig:second!}b and~\cite{Supp}. In addition as expected, we do find that hyperuniformity emerges at the onset of the transition, Fig.~\ref{fig:second!}. The number fluctuations are reduced compared to a random set of points and scale as: $\Delta N^2(\ell)/\langle N(\ell)\rangle\sim \ell^{\lambda^{\infty}}$, where $\lambda^{\infty}~\sim-0.5$ is the steady-state value of  $\lambda$.  Altogether these results confirm that the transition is a  critical phenomenon, which belongs to the same universality class as the discrete  Random-Organization models~\cite{CorteNature2008,Tjhung2016}.
We also note that this behavior is not specific to a choice of $\epsilon$, the phase diagram of systems driven by AC shear flows is provided in Fig.~\ref{fig:second!}e.

At first sight, when looking solely at the fraction of active particles, the transition looks qualitatively similar for a palindromic {\em chaotic} drive, filled symbols in Fig.~\ref{fig:second!}a. However, the similarity stops there. The transition indeed lacks all signatures of criticality. The relaxation time of the order parameter increases but is not seen to diverge at $\phi^{\star}$, Fig.~\ref{fig:second!}b, and more importantly we do not observe any sign of emergent hyperuniformity, see Figs.~\ref{fig:second!}c, \ref{fig:second!}d, and Supplementary Video 3. The number  fluctuations scale as the area of the observation window, indicating an uncorrelated structure at large scales. The difference is even more obvious when looking at the geometry of the phase diagram in the $(\phi,\epsilon)$ plane, Figs.~\ref{fig:second!}e and \ref{fig:second!}f. When the particles are advected by a chaotic flow, no reversible state can be reached above a threshold value, here $\phi~\sim 5\times 10^{-3}$. Even more surprisingly, while $\phi^\star$  decreases with $\epsilon$ for AC shear, it is found to increase with the noise amplitude for chaotic flows: in other words increasing $\epsilon$, the only source of microscopic irreversibility, makes the global dynamics more reversible at a given average area fraction. Back to a mixing perspective, two comments are in order: first the upper bound of $\phi^\star$ is obviously a strong limiting factor. Second and even more importantly,  chaotic periodic driving does not improve at all particle homogenization, rather this strategy yields larger spatial heterogeneities compared to AC shear flows, see Figs.~\ref{fig:second!}c and \ref{fig:second!}d.

\begin{figure}[t]
\begin{center}
  \includegraphics{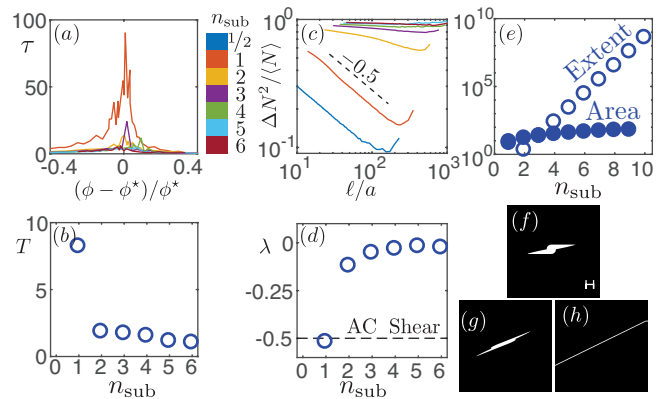}
  \caption{\label{fig:third!}
  (a): Relaxation time $\tau$ at various values of $\nsub$, as $\nsub$ increases criticality is hindered. 
  This trend is also visible in
  (b) where the average relaxation time $T$ is seen to decrease function with $\nsub$.
  (c) As $\nsub$ increases, the power-law exponent $\lambda$ becomes less negative, hence hyperuniformity decreases. The fitted value of $\lambda$ is also plotted in (d).
  (e)The Interaction region elongates exponentially with $\nsub$, causing it to span the full system within a small number of subcycles whereas its area grows only linearly. Here lengths are expressed in units of particle size. (f-h) Examples of interaction domains for (f) $t_x=1$, (g) $t_x=0.5$, $\nsub=1$, and (h) $t_x=0.5$, $\nsub=2$. 
  }
\end{center}
\end{figure}
We now single out the reason for the suppression of criticality by Lagrangian chaos, and introduce optimal mixing strategies around this intrinsic limitation to hyperuniform mixing.
It is illuminating to first address the impact of $\nsub$ both on the  dynamics and structure of the suspension, where, again,  $\nsub$ is the number of alternations between horizontal and vertical shears before reversing the flow.
As a measure of criticality we first compute the average of the relaxation time $T\equiv 2/\phi^{\star}\int \tau (\phi) {\rm d}\phi$  over the domain $|\phi-\phi^\star|/\phi^\star<0.25$.
The results are plotted in Figs.~\ref{fig:third!}a and ~\ref{fig:third!}b. $T$ is highest for $\nsub=1$ where $T\sim 20$ cycles, and drops very sharply to a value as small as 2 cycles cycles as $n_{\rm sub}=2$.  
A very small amount of stretching and folding is clearly  enough to fully suppress critical slowing down. The same trend is observed for the magnitude and spatial correlations of the particle-number fluctuations, Figs.~\ref{fig:third!}c and \ref{fig:third!}d. Close to $\phi^{\star}$ we find that  the exponent $\lambda$ characterizing the scaling of the number fluctuations with the box size, or equivalently the $\lambda^\infty$ characterizing the spatial decay of the pari-correlation function,  is non-trivial only for $n_{\rm sub}\leq2$, Fig.~\ref{fig:third!}d.

We now pinpoint the reason for the absence of spatial correlations for $n_{\rm sub}>2$.  We measure the shape of the region enclosing particles that will undergo contact interaction in the course of one cycle. The size of this region indeed represents the range of the interactions of the strobed dynamics, where the particle positions are observed only at  the  end of each cycle~\cite{CorteNature2008,Schrenk_JChemPhys_2015}. The interaction volume has a $S$-shape in the case of simple shear, Fig.~\ref{fig:third!}f, and its extent scales as $a\gamma_0$, where $a$ is the particle diameter, and $\gamma_0$ the shear amplitude. 
In all simulations, this length is orders of magnitude smaller than the box size. In stark contrast, the essence of chaotic advection is
to exponentially amplify the extent of the volume spanned by a patch of passive tracers~\cite{SouzyJFM2017}. 
Therefore particles belonging to separated patches can come to close proximity after a few folds and stretches. This mechanism sets the effective-interaction volume between the advected particle shown in Figs.~\ref{fig:third!}g and~\ref{fig:third!}h for $n_{\rm sub}=1$ and $2$ respectively. The area of the interaction volume increases linearly with $n_{\rm sub}$, however its largest dimension grows exponentially after each sub-cycle (see also \cite{Supp} where we describe the conditions for exponential growth).  As a consequence for $n_{\rm sub}=10$ and $\gamma_0=3$, as used in the above simulations, the interaction range between particles of $10~\mu\rm m$ 
in size would be of the order of $100$ meters! Needless to say that in any practical applications the interactions would be effectively long-ranged. Nonetheless, the strong anisotropy of the interaction volume allows the system to find absorbing states at finite densities, but both criticality and hyperuniformity are lost.
 The concept of hyperuniform ordering is indeed meaningful only over length scales larger than the particle size, which here turns out to be effectively of the order of the entire box size. In addition, in~\cite{WeijsPRL2015} combining experiments on periodically driven emulsions, and realistic simulations, we also established that long-range hydrodynamic interactions prevent large-scale hyperuniform ordering. Altogether these results strongly suggest that, in contrast with  equilibrium physics~\cite{Lebovitz2000}, long-range interactions are a generic impediment  to hyperuniformity in periodically-driven systems. {\color{black} One final note is in order: Chaotic driving does not only cause the effective size of particles to increase. It also causes interacting particles to exponentially deviate from their unperturbed trajectory, something that does not happen in any discrete random-organization model. The latter effect has  a strong impact on the displacement statistics  of the strobed dynamics as demonstrated in the Supplemental Information~\cite{Supp}.}

\begin{figure}[t]
\begin{center}
\includegraphics[width=\columnwidth]{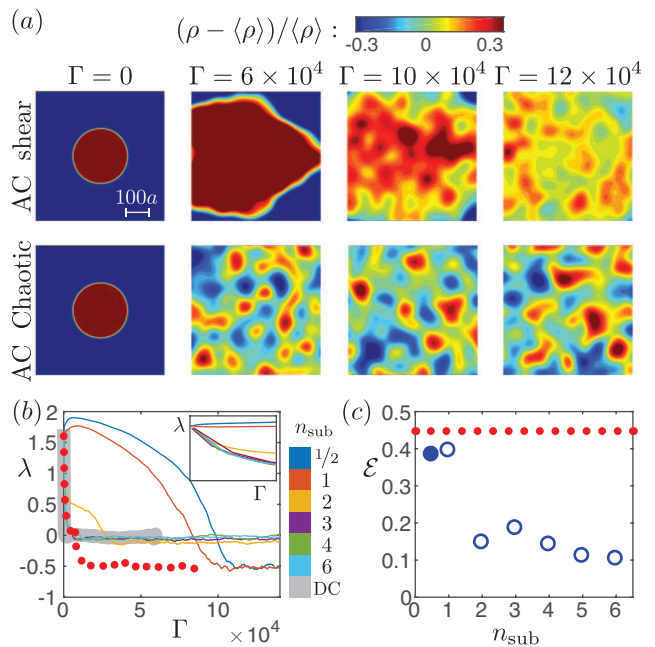}
\caption{\label{fig:four!}
Dispersion dynamics of an initially concentrated patch of particles (initial diameter: $228\,a=0.2$).
{\color{black}  (a): Evolution of the particle-density field with the accumulated strain $\Gamma$ under  shear  (top) and chaotic flows (bottom). {\color{black} Same definition as in Fig. 2}. Both  dispersion and homogeneization are slow  for AC shear, whereas they are very fast for chaotic flows. However these density fluctuations in the steady state are much larger in the chaotic case.}
{\color{black}   (b): Solid lines: Homogeneization (quantified by $\lambda$) as a function of accumulated strain $\Gamma$ for different values of  $n_{\rm sub}$. Only pure shear flow (corresponding here to $\nsub=1/2$) and chaotic driving with $\nsub=1$ lead to a significant degree of hyperuniformity ($\lambda^\infty\sim -0.5$). Both higher values of $\nsub$ and DC chaotic flows lead to fast dispersion, but to normal number fluctuations ($\lambda^\infty\sim 0$). Inset: Close up on the small strain region showing that all the $\lambda(\Gamma)$ collapse on a single master curve above $n_{\rm sub}=3$}. Dotted line: hybrid protocol introduced in the main text. 
   (c)  Plot of $\mathcal E$ as defined in Eq.~\eqref{eq:efficiency} for the various protocols. Filled symbol indicates the pure shear case.
  Dotted line: Efficiency measured for the hybrid protocol introduced in the main text.}
  \end{center}
\end{figure}

We now show how to go around the above limitations by designing sequences of  mixing and unmixing, which optimize both speed and uniformity. From a practical perspective, for a given actuation mechanism, we could be left with three control parameters depending on the applications : the strain amplitude, $\gamma_0$, the particle fraction $\phi$, and the number of flow alternations by stirring cycle $n_{\rm sub}$. However, we  restrain here  to  optimization with respect to $n_{\rm sub}$ only. 
For sake of simplicity, and to facilitate comparison with the  results discussed above, we  focus on a model situation where the distance to the reversible transition is fixed, keeping  $\gamma_0=3$ and adjusting $\phi$.  
Let us now introduce a very natural measure of mixing efficiency $\mathcal E$:
\begin{equation}
\label{eq:efficiency}
{\mathcal E}\equiv1- \frac 1 2\left[(1-|\lambda^{\infty}|)^2+(1-\Gamma_{0}^\infty/\Gamma^{\infty})^2\right],
\end{equation}
{\color{black}  $\Gamma^\infty(n_{\rm sub})$ is the strain accumulated  when steady state is reached, and  
$\Gamma_0^\infty=3000$ is the accumulated strain required to reach steady state in the limit of large $\nsub$. The accumulated strain $\Gamma$ is  defined at the end of each cycle as the product of $\gamma_0$ and the number of oscillations for AC shear, and as the product of $(\gamma_0 n_{\rm sub})$ and  the number of cycles for chaotic drives.} Therefore, $\mathcal E$  is simply defined as one minus the Euclidian distance to the optimal mixing situation 
where the dispersion would be maximally hyperuniform ($\lambda^\infty=-1$), for a minimal  accumulated strain at steady state ($\Gamma^{\infty}=\Gamma_0^\infty$).
So, $\mathcal E=1$ in the optimal case, and $\mathcal E=0$ in the worst case scenario.
We measure this efficiency  starting from an ensemble of particles concentrated inside a circular patch, Fig.~\ref{fig:four!}a. We first plot the variations of $\lambda$ measured at the end of each cycle with the instantaneous accumulated strain $\Gamma$, see Fig.~\ref{fig:four!}b and Supplementary Video 4~\cite{Supp}.  We clearly see that  $\lambda$ undergoes a fast monotonic decay toward values of $\lambda^{\infty}$ close to $0$ for $n_{\rm sub}>2$. 
For AC shear flows labeled at $\nsub=1/2$ in Fig.~\ref{fig:four!}, these variations are  slow but converge to  $\lambda^\infty\sim -\frac 1 2$. These opposite trends are reflected by Fig.~\ref{fig:four!}c where $n_{\rm sub}=1$ is found to optimize mixing by unstirring. 
An even more efficient strategy can be provided if the actuation can be switched in the course of the dispersion. The advantage of the different types of flows flows can then be sequentially exploited. Having a double-stirrer geometry in mind (\eg that of an egg beater) the advection can be easily switched from simple to chaotic by either driving the stirrers synchronously or asynchronously~\cite{aref1984}.  
An idealized example is illustrated in Figs.~\ref{fig:four!}b and ~\ref{fig:four!}c, where the (red) dotted lines correspond to a sequence where first normal chaotic advection is used to quickly disperse the particle cloud, followed by a second sequence where an AC shear flow close to the reversible transition organizes the particles in a hyperuniform structure. The efficiency of this combined stirring protocol exceeds that of any pure shear or chaotic palindromic sequence Figs.~\ref{fig:four!}c.

In conclusion, we have explained how chaotic advection  hinders the emergence of hyperuniform ordering upon periodic strirring. Introducing a natural measure of mixing efficiency, we have proposed effective  strategies combining both fast spreading and hyperuniform dispersions out of reach of conventional mixing protocols. 

\acknowledgements
We thank R. Jeanneret, R. Dreyfus, and E. Villermaux for valuable comments and suggestions.
We acknowledge support from Institut Universitaire de France (D. B.), and the NWO Rubicon programme financed by the Netherlands Organisation for Scientific Research (J. H. W.).

%

\end{document}